# Reducing the uncertainty in the forest volume-to-biomass relationship built from limited field plots

Caixia Liu[1], Xiaolu Zhou[2*], Xiangdong Lei[3], Huabing Huang[1*], Changhui Peng[2], Xiaoyi Wang[1], Jianfeng Sun[4], Carl Zhou[5]

[1]State Key Laboratory of Remote Sensing Science, Jointly Sponsored by the Institute of Remote Sensing and Digital Earth of Chinese Academy of Sciences and Beijing Normal University, Beijing, 100101, China;

[2]Ecological Modeling and Carbon Science, Department of Biology Science, University of Quebec at Montreal, Montreal, QC H3C 3P8, Canada;

[3]Institute of Forest Resource Information Techniques, Chinese Academy of Forestry, Beijing 100091, China;

[4]Guangzhou University, School of Geographical Science, Guangzhou 510006, China;

[5]Faculty of Health Sciences, University of Ottawa, Ottawa, ON K1N 6N5, Canada.

**Summary**

1. The method of biomass estimation based on a volume-to-biomass relationship has been applied in estimating forest biomass conventionally through the mean volume ($m^3$ $ha^{-1}$). However, few studies have been reported concerning the verification of the volume-biomass equations regressed using field data. The possible bias may result from the volume measurements and extrapolations from sample plots to stands or a unit area. This paper addresses (i) how to verify the volume-biomass equations, and (ii) how to reduce the bias while building these equations.

2. The volume-biomass equations are re-expressed by two parametric equations. The stem biomass plays the parameter's role in the parametric equations, of which one is established by regressing the relationship between stem biomass and volume, and the other is created by regressing the stem allometric equation. The graphical representation of parametric equations proposes a concept of "restricted zone", which helps to verify the volume-biomass equations in regression analyses of field data.

3. The regressions of the parametric equations are verified for 30 species and forest types corresponding to the three datasets. This study demonstrates that most species, widely distributed in cold, temperate, and subtropical zones, is consistent with the allometric curves. The wood density impacts parameters of volume-biomass equations more than the stem allometric curve does. Our analyses suggest that the averaged allometric equations and specified wood density could be an alternative solution for building volume-biomass equations for those species not yet having field biomass measurements. All these analyses and calculations are based on field measurements from three large datasets for both volume and forest biomass (total 10,840 stands), and one for wood density (16,468 estimates).

4．This paper presents an applicable method for verifying the field data using reasonable wood densities, restricting the error in field data processing based on limited field plots, and achieving a better understanding of the uncertainty in building those equations. The verified and improved volume-biomass equations are more reliable and will help to estimate forest carbon sequestration and carbon balance at any large scale.

**Introduction**

Various techniques have been developed for observing today's natural world. Unfortunately, the measurements of large-scale forest biomass cannot be conducted directly due to restrictions like heavy load of fieldwork, non-destructive measurement requirements, etc. Indirect methods have been applied in estimating forest biomass conventionally through the amount of growing volume, i.e. the method of biomass estimation based on volume-to-biomass relationship (Brown & Loge 1984; Brown 1997; Somogyi et al., 2007). However, these estimates of forest biomass may frequently raise an issue of accuracy. Although the accurate volume of forests can be measured by sampling under certain precision at large scale, the error that occurs in the transformation from volume to biomass has been an issue for decades, especially at global and national scales. Studies have discussed that the estimation of a forest's biomass and its carbon sequestration capacity on a national scale have often been carried out with great uncertainty (Goodale, Quere & Raupacha 2002; IPCC, 2013; Houghton, 2005). This uncertainty could be implicit in the relationship of volume-to-biomass for most tree species. The forest biomass estimates with significant error, if any, will influence our understanding of global carbon balance.

An error can occur from different procedures of a whole estimation of forest biomass for many reasons. The most essential reason is the invisibility of the tree parts underground, which is usually around 20% of total biomass for coarse roots and over 10% for fine roots. That is why harvesting whole trees, including roots, in destructive plots (ranging in size from 100 to 3000 $m^2$) is so important for creating volume-to-biomass relationships. These relationships are often modelled using regression equations with total biomass (t ha$^{-1}$) as dependent variable, and stem volume (m$^3$ ha$^{-1}$) as independent variable. Many of these equations have been developed for specific species or forest types since the last century. For hundreds of these equations, the uncertainties are not even capable of being assessed for representative populations, let alone being eliminated. The volume-biomass equation is somewhat similar to a black-box between volume input and biomass output. Although numerous studies have reported on techniques for making volume-biomass equations, including sampling, measurement, and regression (see summaries and discussions presented by Parresol, 1999; Smith, Heath & Jenkens 2002; Jenkins, Chojnacky & Heath 2003; Zianis, Muukkonen & Makipaa 2005; Picard, Saint-André & Henry 2012; Jucker, et al. 2016; Yuen et al., 2016), we are still facing two tough questions for reducing the uncertainty: (i) What field plot size is appropriate? (ii) How many field plots are sufficient? For the former, the problem is that plots that are too small cause counting errors on the statistics in a unit area, which makes regressions that differ widely. For the latter, too few plots cannot represent the population of forest stands. Nonetheless, building larger and more plots is unachievable due to measurement cost. This has created an urgency to find a method that ensures accuracy in describing volume-to-biomass relationships with lower field work costs.

  This paper discusses equation forms of volume-biomass based on measurements at stand level, and introduces a new form as parametric equations that conduct stem biomass for an intermediate variable. Our algorithm extracts the information through a different regression method of field data. This information is usually ignored but can be used to limit the range of regression error, for example, to restrict the error by an easily measured variable, such as wood basic density (WBD, a term in wood science) that can be found from timber industry publications. The presented analysis framework focuses on a new understanding of the information from limited field plots, and suggests a new approach for assessing the regressions of volume-biomass.

Notice that this study does not discuss any other biomass equations containing those variables, such as, diameter at breast height (D), tree height (H), and their combined forms (DH or $D^2H$).

## Materials and methods

### STRATEGY

For the purpose of converting forest volume to total living forest biomass, a widely used biomass function with an independent volume variable is $B_t = \alpha V^\beta$ ($B_t$ is total biomass including foliage, branch, stem, and root biomass, t ha$^{-1}$; V is stem volume, m$^3$ ha$^{-1}$; $\alpha$ and $\beta$ are parameters). As mentioned previously, to determine this equation requires a number of field plots, and most likely yields a markedly bias if the plots are insufficient to determine $\alpha$ and $\beta$ by regressing. This is because these types of regressions are normally parameterized by regressing V on $B_t$ directly instead of analyzing model behavior. That is to say, the conventional method focuses on results rather than causes. Since there is no apparent physiological relationship between volume and total biomass, then whether or not to create a direct causal relationship of V and $B_t$ becomes an issue of interest. Generally, this issue can be abstracted as a black-box problem, which should be solved gradually in stages to reduce its uncertainty (Fig. 1). The uncertainty can be identified according to the general technical flow of nonlinear system identification, namely, data preparation, model postulation, parameter identification, and model validation. Such a procedure makes it clear that the traditional volume-biomass equation needs to be converted to a set of parametric equations. These postulated equations and related algorithm are explained in the following sections in detail.

**PARAMETRIC FUNCTIONS**

In applications of function and equation, the parametric equation is mainly utilized to solve problems in multidimensional space for convenience of mathematical treatment. Beyond this, it may also express some physical or physiological quantity to clearly and more effectively describe the relationship between the parameter and its function. Here, the parameter (parameter variable) is defined as the independent variable in a set of parametric equations. This special relationship would be the key that we want to pay more attention to in our analysis. Hence, we introduce a parameter variable $B_s$ (stem biomass, t ha$^{-1}$) and make a pair of equations to separate the error source: $B_t = aB_s^b$ and $V = B_s/\rho$. For general expression, we give the following parametric equations,

$$\begin{cases} B_t = aB_s^b & \text{eqn 1} \\ B_s = \rho V & \text{eqn 2} \end{cases}$$

where $B_t$ denotes total biomass (t ha$^{-1}$), V is mean volume (m$^3$ ha$^{-1}$), ρ expresses wood density (t m$^{-3}$), and a and b are parameters of the allometric function. Eqn 1 actually expresses the inverse allometric function of $B_s$, and still reflects the allometric relationship between $B_t$ and $B_s$.

Both eqn 1 and eqn 2 reflect tree physiological characteristics. Eqn 1 expresses the changing ratio of stem biomass to total biomass; eqn 2 indicates the linear relationship of stem biomass and volume by a coefficient ρ. The undetermined parameters a and b are solvable since field measurements of both stem biomass and total biomass are available. While parameterizing the two equations, we can examine whether the parameters are reasonable and reliable depending on our tree physiological and wood physical knowledge. If all parameters (a, b, and ρ) are reasonable, there must be a "restricted zone" existing between the curves of the two equations with the assumption that (1) a stem is lighter than the whole tree, and (2) wood density is generally less than 1.0 (t m$^{-3}$). This restricted zone is represented in Fig. 2-B as an example. Any observation lying into the restricted zone strongly suggests that there are problems in field measurement or counting. After fixing these problems by setting up restrictions on regression parameters, the reliability of these two parametric equations will be improved and eqn 1 and 2 can then be rewritten as a conventional single equation:

$$B_t = \alpha V^\beta \qquad \text{eqn 3}$$

where $\alpha = a\rho^b$, $\beta = b$, and ρ is wood density (t m$^{-3}$), and both $B_t$ and V are the values per unit area.

**PARAMETER IMPROVEMENT**

To follow the above strategy and formulation, three points should be noted: (i) Improving parameters by utilizing parametric functions is a "separate-to-recombine" process (Fig. 2), in which we can find that the power $\beta$ is the net result of allometric measurements on the stem. This is because the curvature of eqn 3 is only affected by $B_s$-to-$B_t$ allometric relationships, but not by volume V. If directly regressing field data for the variable $B_t$ and V, the power $\beta$ will usually be impacted more or less by possible outliers of V. Outliers make wood density estimates largely inconsistent, which changes the curvature (i.e. $\beta$) of eqn 3. In short, the $\alpha$ and $\beta$ will change after carrying out separate-to-recombine processing that forces $\rho$ to be unique for a specified species. (ii) Through the separate-to-recombine procedure, the impacts of uncertainty or mistake in volume measurement can be completely excluded from parameter $\beta$. We may only focus on improving the parameter $\alpha$ by examining $\rho$. This is why we process the relationships of $B_s$-to-$B_t$ and V-to-$B_s$ separately. This meets the needs of the model postulate to divide the "black-box" into two parts based on the concept of system identification. (iii) The $\rho$ can be corrected by comparing with the WBD, although $\rho$ might be lower than it since WBD does not reflect bark biomass. Thus, possible significant biases are avoided for the parameters in eqn 3.

Summarily, we first separate the equation into two parametric equations (Fig. 2-A). Secondly, carry out regressions respectively for the two equations (Fig. 2-B). Thirdly, find possible observational errors from the V-to-$B_s$ equation according to make a decision to improve the parameter ρ for the equation, and finally combine the two equations back into one standard allometric equation as was the original form (Fig. 2-C). During this procedure, it must be emphasized that the restricted zone is variable depending on decisions for different species. The Global Wood Density Database (Zanne et al., 2009) gives that wood density ranges from 0.2 to 1.3 for most species. This dataset and WBD from timber industry information are our knowledge to improve ρ.

**DATA DESCRIPTION**

We used three large datasets based on over 10,840 measurements of field plots, which were collected from a large data compilation from eastern European (Usoltsev, 2013; 8,033 plots), Chinese (Luo, Zhang & Wang 2014; 1,607 plots), and western European and Japanese scientific literatures (Cannell, 1982; over 1,200 plots). The dataset includes growing stock volume (V), total biomass ($B_t$) and stem biomass ($B_s$). The biomasses in the datasets were usually measured based on harvests of subsets of trees. At some plots, growing stock volumes (V) are not directly provided in the datasets, and were calculated based on the empirical formulae, which includes variables of stand mean of diameter at breast height (DBH), height, and tree density if they were available (Luo et al., 2014). Since data from Luo et al. (2014) are the latest collection for China's forests, we excluded all records of measurements in China from the Usoltsev's dataset.

Of the total 10,840 sample plots assembled in the three datasets, the applicable data (9,227

records) processed in this study correspond to 3,335 plots for total biomass and stem biomass, and 3,399 plots for growing stock volume and stem biomass. For Luo et al. (2014) and Usoltsev (2013) the trees were basically grouped by species, while for Cannel (1982) all trees were grouped by forest types (conifer, broadleaved, mixed and tropical forest). All these data were complied as stand level by data source providers, such as stand age (year), mean volume ($m^3 h^{-1}$), and mean biomasses (t $ha^{-1}$). We also employed a large compilation of global wood density data (16,468 records) encompassing 8414 taxa, 1683 genera, and 191 families from Zanne et al., (2009).

**COVERAGE AREA OF OBSERVATIONS**

The spatial locations of all field plots in the datasets are widely distributed geographically in 48 countries around the world (Fig. 3). The measurements were carried out with or without belowground biomass for over 317 tree species. The dataset includes different forest types at different latitudes and climatic conditions. These plots are disbursed over boreal temperate, subtropical, and tropical zones with forest ages from young stands of about 5 years to over mature forest of more than 400 years. Most types of woody plant stands are represented including those from natural and plantation forest origins, ranging from oak woodlands and coniferous plantations to tropical rainforests and mangrove swamps. The distribution of plots is uneven across continents, in which they are highly concentrated in some countries (Fig. 3).

## Results

### COMPARISON OF THE RELATIONSHIPS

To avoid possible statistical error caused from original compiles of the dataset, we respectively analyzed three independent data sources published in different times (refer to Table 1). Our results report that the relationship between $B_s$ and $B_t$ have a better fitting performance than V and $B_s$ (Fig. 4). Most coefficients of determination for $B_s$ vs $B_t$ are greater than 0.9. In the bottom right part of each scatterplot, $B_s$ and V have a significant linear relationship with all coefficients of determination greater than 0.6, although a few samples are located in the restricted zones for some species and forest types (No. 8, 10, 15, 18, 25, 27, 29, 30). All regression curves ($B_s$-to-$B_t$) and lines (V-to-$B_s$) avoid the "restricted zones".

The modeling parameters (*a*, *b*, and *ρ*) are listed in Table 1 for 30 species and forest types. The values of *b* are distributed in the vicinity of 0.9 and the values of *a* have a wide range from 1.88 (No. 25) to 4.5 (No. 6). As for the estimated *ρ*, which is the slope of the regressed straight-lines in Fig. 4, all values are less than 0.7, and these should not be higher than 1.0 (t $m^{-3}$). Tropical trees have a relatively higher wood density with values greater than 0.6.

### PARAMETERIZATION OF VOLUME-BIOMASS EQUATIONS

After determining *a*, *b* and *ρ*, the parameter *α* in volume-biomass equation (eqn 3) can be calculated as $\alpha = a\rho^b$. Based on the dataset of wood density measurements compiled by Zanne et al., (2009), we selected measured wood densities to match the species and types classified in Table 1. These wood densities ($1^{st}\rho$ in Table 1) are the slopes of regression equations in Fig. 4, The values of $1^{st}\rho$ are quiet different from WBD ($2^{nd}\rho$ in Table 1) for some species. The percentage errors on $\rho$ range from -16% to 18% (average 8%; Table 1). The comparison between the values of the parameter $\alpha$ calculated based on different values of $\rho$, is presented in Table 1 in order to understand how much $\rho$ could influence $\alpha$. Our calculations show that the $\alpha$ values differ partly because they are affected by $\rho$ slightly due to the differences between regressed and measured wood densities.

**Discussions**

**REGRESSION MODEL TEST**

There are two common ways to test a model and evaluate if it is better than others. A widely used method is the comparison between prediction and independent observation. Unfortunately, we do not have other field data independent of the measurements used in this study. Another way is statistical hypothesis testing, which rejects or fails to reject (does not equal accept) null hypothesis. However, the significance level or p-value provided by a parameter test cannot prove that a model result is closer to the true value than other models. For the immeasurable forest population, model test is always difficult when we are facing a large-scale forest. Given this consideration, we made a dummy population to realize the true values of a large-scale forest. This population consists of 10,000 forest stands at the regional scale based on the output of pseudorandom number generator (Table S1, Fig. S1). The true values of total biomass amount are counted by enumerating all dummy stands in the population for model comparisons. Our experiment consists of 500 random samplings (N=500), which simulates the behaviors of establishing a few temporary sample plots (or collecting field data) to monitor the population 500 times.

The results of parameter verification indicate two major points: (1) If all stands have the same features (wood density and ratio of stem to total biomass) and have no measurement errors either, it makes no difference (Table 2) whether the equation is improved or not via the separate-to-recombine processing. (2) As long as the stand features become dispersible for realistic forest, and the errors occur in practical measurements, the equation improvement results in different parameters. The improved volume-biomass equations are better than the ones before improvement, based on choosing both 20 and 100 plots per sampling (n=20 and 100, Fig. 5).

The comparison illustrates that 100-plot-sampling parameterizes the equation well. The simulations exhibit different probability distributions of the parameters ($\alpha$ and $\beta$) and residuals (e), of which improved $\alpha$, $\beta$ and e generally have low standard deviations (Table 2). The convergence rates are faster on 100-plot-sampling than 20-plot-sampling for all tests of $\alpha$, $\beta$ and e (Fig. 5). Nonetheless, differing from the stochastic simulation, in practice, only one sampling can be implemented in field measurements (or data collection). Having such few data makes us, obviously, prefer the improved volume-biomass equations so as to cause less residual errors (it is decreased by up to 50% in Table 2) with high probabilities than the original equations.

To understand the differences of total biomass amounts affected by variant wood densities, we estimated the biomass of China's *Eucalyptus* forests, which are scattered in 10 provinces over a large area (total 4455.2 K ha) (CMF, 2013). The five estimated wood densities $\rho$ (listed in Fig. 2) result in large gaps (Fig. 6). This implies the importance to estimate wood density correctly. If using the average (0.54 t $m^{-3}$) of two values from biomass measurements (0.46 t $m^{-3}$) and WBD (0.61 t $m^{-3}$), the total biomass amount lightly decreases to 1.66 (M t) from 1.69 (M t) (Fig. 6) after improving the parameters ($\alpha$, $\beta$).

**UNCERTAINTIES**

Our experiment and estimation based on both dummy and practical data indicate that once mean volumes ($m^3$ $ha^{-1}$) are known for each species in a large area, the equations that transform volume to biomass primarily affect the extent of uncertainty in the biomass estimation of a forest ecosystem. This means that the uncertainties can be seen as being caused by the volume-biomass equations. If searching for the source of measurements, we can find that uncertainties may result from the following derivation: eqn 3 is always used as a function for the mean values of biomass (t $ha^{-1}$) and volume ($m^{-3}$ $ha^{-1}$); these reflect that the values are averages of all measurements at stand level; a stand mean value is usually calculated by extrapolating from plot(s), namely derived from a few individual trees. We can deduce that these inevitably impact observed volume-to-biomass relationship and produce errors (Fig. 2-A), which may involve issues such as whether or not sampling is representative, the acceptance or rejection of diameter class, mistakes in counting or accumulating, etc. Additionally, any operational error is also extrapolated to stand level and will be introduced into the volume-biomass equations. Technically, it is easier to measure each organ biomass (e.g., foliage, branch, stem, and coarse and fine root) of a few trees accurately than to measure the mean volume ($m^{-3}\,ha^{-1}$) of a stand. The volume calculation may cause statistical errors due to the heterogeneity of each stem form and diameter class.

Our analysis revealed the fact that the measurement of volume causes uncertainty of regression parameters more than the measurements of stem and other organs do (e.g. leaf, branch, fine root, and coarse root). The data marks (Fig. 2-B and Fig. 4) invaded the "restricted zone" more frequently in the bottom right rather than the top left of the figure. The observed stem biomass suggests a relatively large and stable portion of total biomass regardless of a single tree or plots. Based on the dataset presented by Usoltsev (2013), consisting of measurements for both stem and total biomass from 1,936 plots in total, the percentage of stem in total biomass is averaged as 67.96% (SD=7.83%, n=1,502) for all plots where the ages were less than 30 years. In contrast, ratios of stem biomass to volume were spread over a wide range. For example, the deviation (see Fig. 7) is higher in the bottom right zone (SD=22.4% for non-tropical and 25.1% for tropical forest) than the top left zone (SD=16.6% for non-tropical and 17.5% for tropical forest). The coefficients of determination (CD) also support this finding, and report that all CDs (see Fig. 4) are higher for eqn 1 than eqn 2 without exception. This may in fact reflect where the major observation uncertainty comes from. That is particularly why the measurements need to be inspected for the mean volume ($m^{-3}\,ha^{-1}$).

Fig. 2-B illustrates that the data plotted in the bottom right zone (green area) exposed some abnormal dots located in the restricted zone (red area). Some dots overstepped the boundary; some are located near the horizontal axis. The difficulty arises when judging whether the average volume density is too low or too high, because of a large discrepancy between proportions ($B_s/V$) and reasonable corresponding wood densities. These errors were hidden in the pre-separation volume-biomass equation (Fig. 2-A). Therefore, our analysis suggests that it is capable of wood density acting as an inspector to verify the rationality of field measurements. Arguably, as long as we are certain about the relationship of stem and total biomass (eqn 1), any existing volume-biomass equation with a power function form can be tested to see whether the measurements deviate too far from reasonable wood densities or not.

**FIELD WORKLOAD**

The measuring of forest biomass has been, at least so far, a difficult experience in terms of field workload. People's measuring ability is always limited. Due to such limitations, the requirements of big plots can rarely be met. Though the size of field plots for volume measurement can expand to an entire stand for complete enumeration, tree biomass is conventionally measured destructively on several sample trees. As for the measurement of underground biomass, it is limited more by a measurer's operability. The usual way is to harvest a few trees including all parts (except soil coring for fine roots) and to use the information of their total biomass and volume as being representative of the volume-to-biomass relationship for one unit of forested area. The issue that we are facing is whether by only harvesting a few trees it is possible to sufficiently represent an ecosystem's properties. Our analysis suggests that parametric equations would be a better means to lower the field workload.

First, since eqn 1 only express the allometric relationship of a stem organ and the whole individual, this relationship is not affected by volume regardless of single tree or stand mean. It implies that the stem ratio of total biomass ($B_s/B_t$) observed by harvesting a few trees should be consistent with the tree population of a species. Moreover, our analysis found that this ratio is almost the same even for some different species. For example, a small range (within +/-5%, Table S2) denotes a slight difference of average ratios between each species and four selected species (Nos. 1, 2, 3, and 4). It implies that quite a few species have a close $B_s/B_t$ ratio. In other words, the parameters (a and b) in eqn 1 are relatively stable in different measurements. Hence, it is possible to estimate the $B_s/B_t$ ratio of a population by using a few trees under normal growing conditions. Therefore, a field measurement should focus on the allometric relationship rather than the organ biomass of stand mean, if the field data are only used for volume-biomass equations. This has practical significance for reducing field workload.

Second, making a connection between volume and stem biomass facilitates parameterizing eqn

2. These two variables can essentially determine wood density, and accordingly field measurements might bring about less tree damage by collecting wood samples rather than conducting a destructive measurement of plot biomass (e.g. cutting down all trees). In addition, comparing the slope of the equation with WBD (bark effects removed) will help us to judge if measured values are located in a right range, and then help us to recognize outliers early to save time in field work. The practical measurements require an important tenet to limit the $\rho$ within a minimum range. This can effectively exclude possible illogical results.

From the above perspective, it is not necessarily required to set up large, or many, field plots for understanding the volume-to-biomass relationship. This will cut costs in the field biomass measurement of forest ecosystems.

## BIOMASS ESTIMATION ON LARGE SCALE

Most forest ecosystems are comprised of various tree species; we cannot realistically create volume-biomass equations for each species. The lack of equations makes processing inconvenient for species classification, and makes matching every species with a proper equation very confusing. If this is the case, those species could be modeled and parameterized by using the averaged $B_s/B_t$ ratio and existing wood densities. Our analysis found that most species that are distributed widely in both cold temperate and subtropical zones exhibit good consistency of the allometric curves (see $B_{t(B_s)}$ in Fig. 4). The curves only show slight differences between conifer and deciduous trees. The averaged allometric curves are illustrated in Fig. 7, which summarizes valid data collected from 3399 field plots across the world and suggests two sets of parameters in eqn 1 (i.e. the allometric curve) applicable to non-tropical and tropical zones.

If adopting an allometric curve for a species, its volume-biomass equation only depends on the wood density ρ, which is obtained from either field biomass measurements or WBD. The biomass estimated using WBD should not show considerable differences, although WBD disregards bark effects. Practically, the proportions of bark volume and quality compared to stem are quite close. We reviewed the literature discussing proportions of bark volume (Smith & Kozak 1971; Nyg & Elfving 2000; Miles & Smith 2009; Wehenkel, Cruz-Cobos & Carrillo 2012; Murphy & Cown 2015), and calculated the percentage of bark biomass based on field data compiled in the datasets (Connell, 1982; Usoltsev, 2013; Luo, Zhang & Wang 2014), which have bark measurements of 3,211 plots in total. The difference between proportions of bark volume and biomass is not more than 3% on average; therefore the wood densities ρ obtained from calculating with or without the bark do not vary considerably. This means that WBDs offered by the timber industry information can be employed directly.

In building a volume-biomass equation using a set of parametric equations, we noticed that the wood density impacts the equation parameters more than the stem allometric curve does. As seen in Fig. 7, the dispersions between the regression curves and data dots are higher for $B_{s(V)}$ (SD=25.1 and 22.4 for eqn 2) than $B_{t(B_s)}$ (SD =17.5 and 16.6 for eqn 1). That is, the average allometric curves would be more representative of the populations. Considering the above, utilizing the averaged allometric equations and specified wood density could be an alternative solution for building volume-biomass equations for those species not yet having field biomass measurements. Furthermore, for all species, there are "high probability" regions around the typical $B_s/B_t$ ratio (0.68), and "low probability" regions as we know ρ should not be in that area. If the average ρ of a species or region does not correspond with our knowledge, further study is needed to combine prior knowledge about Bs/Bt ratios and wood densities so that each observed point can be checked whether it receives an estimate of the probability of that being a real observation or the probability of that being a measurement or counting error.

**Conclusion**

The uncertainty in the relationship of forest volume-to-biomass can be considered to come from two aspects. One is that limited field plots (for the measurements of both total biomass and volume) may not represent the population well; another is that the volume measurement may easily introduce bias when estimating the mean volume owing to multiple causes. The parametric equation method is a convenient and realistic tool for reducing uncertainty in the relationship of forest volume-to-biomass based on limited field plots. The graphical representation of parametric equations proposes a concept of "restricted zone", which helps to verify the volume-to-biomass relationship in regression analyses of field data. Obeying the limits of a "restricted zone", the knowledge of wood densities can be an inspector for checking field data, and even be an alternative for measurement or extrapolation of the mean stem volume per unit area. The presented analyses of formulating volume-biomass equations suggest an applicable method for restricting the error in field data processing, and achieving a better understanding of the uncertainty in building those equations. The verified volume-biomass equations will hopefully be able to play a significant part in estimations of forest carbon sequestration and carbon balance at any large scale.

**Acknowledgement**

We thank Vladimir A. Usoltsev for providing all files of data in the publication Biomass and Primary Production of Eurasian Forests. This work was supported by Open Fund of State Key Laboratory of Remote Sensing Science (OFSLRSS201404) and Meteorology Scientific Research Fund in the Public Welfare of China [GYHY201506010].

.

**Data accessibility**

The data used in this study include: (i) the field data of biomass and volume across world: Supporting information 2 (Usoltsev, 2013), Supporting information 3 (Cannell, 1982), and forest biomass measurements in China (Luo et al., 2013 and 2014) (http://onlinelibrary.wiley.com/doi/10.1111/2041-210X.12505/full); (ii) wood density measurements across world (Zanne et al., 2009) (http://datadryad.org/handle/10255/dryad.235).

**Authorship**

XZ and CL conceived this study, performed mathematical analysis, and wrote the manuscript; CL, XL, HH, CP and XW discussed the algorithm and its explanations; JS and CZ conducted the data collection, data analysis, and manuscript preparation.

**Supporting information 1**

Table S1 and Fig. S1 (The simulated dataset), Table S2 (Comparison for the ratios of stem to total biomass)

**Supporting information 2**

Biomass measurements (Usoltsev, 2013)

**Supporting information 3**

Biomass measurements (Cannell, 1982)

Table 1. Parameters in eqn 1, 2, and 3 for 30 tree species and forest types across the world

| No. | Species or types* | Based on biomass measurements[‡] | | | | Based on wood density measurements[§] | | | | Error on $\rho$[‖] |
|---|---|---|---|---|---|---|---|---|---|---|
| | | $a$[†] | $b$[†] | 1st $\rho$ | $\alpha$ | 2nd $\rho$ | Range | $\sigma$ | $\alpha$ | |
| 1 | *Abies* and *Picea* | 3.32 | 0.86 | 0.41 | 2.29 | 0.35 | 0.27-0.44 | 0.06 | 1.35 | 18% |
| 2 | *Cupressus* | 3.20 | 0.85 | 0.42 | 2.19 | 0.46 | 0.39-0.51 | 0.06 | 1.66 | -9% |
| 3 | *Larix* | 1.91 | 0.95 | 0.45 | 1.89 | 0.47 | 0.39-0.60 | 0.07 | 0.93 | -5% |
| 4 | *Pinus tabulaeformis* | 2.90 | 0.86 | 0.47 | 1.69 | 0.39 | 0.37-0.41 | 0.02 | 1.29 | 21% |
| 5 | *Pinus koraiensis* and other temperate pines | 3.40 | 0.84 | 0.39 | 2.18 | 0.40 | 0.37-0.45 | 0.03 | 1.58 | -2% |
| 6 | *Pinus yunnanensis* and other subtropical pines | 4.50 | 0.80 | 0.43 | 4.04 | 0.44 | 0.36-0.51 | 0.07 | 2.34 | -1% |
| 7 | *Cunninghamia lanceolata* | 2.52 | 0.89 | 0.35 | 2.16 | 0.31 | 0.26-0.36 | 0.03 | 0.89 | 13% |
| 8 | *Pinus massoniana* | 1.96 | 0.93 | 0.47 | 1.49 | 0.48 | 0.43-0.56 | 0.05 | 0.99 | -2% |
| 9 | Other conifer trees | 3.80 | 0.83 | 0.35 | 1.47 | 0.31 | 0.28-0.39 | 0.04 | 1.44 | 12% |
| 10 | Oaks and other deciduous trees | 3.15 | 0.87 | 0.67 | 2.00 | 0.65 | 0.42-0.80 | 0.11 | 2.16 | 3% |
| 11 | *Populus* and *Betula* | 2.07 | 0.92 | 0.41 | 1.79 | 0.45 | 0.32-0.55 | 0.09 | 0.99 | -8% |
| 12 | *Eucalyptus* and other fast-growing trees | 2.85 | 0.87 | 0.56 | 1.44 | 0.61 | 0.49-0.83 | 0.13 | 1.86 | -7% |
| 13 | Soft broadleaved trees | 2.78 | 0.87 | 0.40 | 1.97 | 0.36 | 0.24-0.49 | 0.12 | 1.14 | 12% |
| 14 | Mixed conifer and deciduous forests | 4.28 | 0.80 | 0.41 | 2.17 | 0.49 | 0.26-0.79 | 0.12 | 2.42 | -16% |
| 15 | Other hard broadleaved trees | 4.27 | 0.80 | 0.48 | 1.29 | 0.49 | 0.24-0.80 | 0.11 | 2.41 | 9% |
| 16 | *Pinus* | 2.57 | 0.89 | 0.44 | 1.25 | 0.40 | 0.37-0.45 | 0.02 | 1.14 | 11% |
| 17 | *Abies* and *Picea* | 2.98 | 0.87 | 0.41 | 1.35 | 0.35 | 0.32-0.37 | 0.02 | 1.19 | 17% |
| 18 | *Fagus, Acer, Carpinus* and *Quercus* | 2.90 | 0.88 | 0.54 | 1.73 | 0.60 | 0.51-0.82 | 0.06 | 1.86 | -9% |
| 19 | *Betula* | 3.14 | 0.85 | 0.51 | 2.06 | 0.54 | 0.50-0.60 | 0.04 | 1.86 | -5% |
| 20 | *Larix* | 2.22 | 0.91 | 0.47 | 1.07 | 0.46 | 0.45-0.49 | 0.02 | 1.10 | 2% |
| 21 | *Alnus* and *Populus* | 2.13 | 0.91 | 0.44 | 1.06 | 0.40 | 0.35-0.44 | 0.04 | 0.92 | 10% |
| 22 | *Tilia* | 3.42 | 0.84 | 0.44 | 1.83 | 0.40 | 0.36-0.42 | 0.03 | 1.59 | 10% |
| 23 | *Castanopsis, Cryptomeria,* and *Pseudotsuga* | 3.39 | 0.84 | 0.39 | 1.59 | 0.41 | 0.31-0.46 | 0.05 | 1.60 | -6% |
| 24 | *Chamaecyparis obtusa* | 3.23 | 0.86 | 0.43 | 1.56 | 0.43 | 0.31-0.50 | 0.06 | 1.57 | -1% |
| 25 | Eucalyptus and other fast-growing trees | 1.88 | 0.96 | 0.64 | 1.22 | 0.63 | 0.49-0.97 | 0.1 | 1.21 | 2% |
| 26 | Conifer | 2.59 | 0.90 | 0.41 | 1.17 | 0.41 | 0.28-0.59 | 0.07 | 1.16 | -1% |
| 27 | Broadleaved | 2.90 | 0.88 | 0.53 | 1.61 | 0.58 | 0.28-0.91 | 0.12 | 1.80 | -9% |
| 28 | Mixed | 4.04 | 0.81 | 0.40 | 1.95 | 0.47 | 0.35-0.62 | 0.10 | 2.19 | -15% |
| 29 | Tropical | 3.71 | 0.86 | 0.65 | 2.56 | 0.61 | 0.39-1.06 | 0.17 | 2.43 | 7% |
| 30 | Tropical | 3.30 | 0.87 | 0.61 | 2.43 | 0.61 | 0.39-1.06 | 0.17 | 2.15 | 0% |

*These species or types were collected by Luo et. al. (2014) (No. 1-15), Usoltsev (2013) (No. 16-25), Connell (1982) (No. 26-29), and Brown & Lugo (1984) and Luo et. al. (2014) (No. 30) from tropical life zones including dry, wet, lower montane rain, moist, montane wet, and premontane forests. The forest types are summarized as conifer, broadleaved, and mixed forest. †The a and b are obtained by regression; b is same as β in eqn 3. ‡The ρ is the slope of regressed straight-lines in Fig. 4; the parameter α is decided by a, b, and ρ ($\alpha = a\rho^b$). §The averaged wood density ρ (t m$^{-3}$) and mean square deviation σ (standard deviation, t m$^{-3}$) are calculated based on a global wood density dataset complied (Zanne et al., 2009). ‖The error on ρ is calculated as (1st ρ – 2nd ρ)/(2nd ρ) *100%.

Table 2. The comparison between the results before and after improving parameters (α and β) of volume-biomass equations. These equations are regressed based on the simulation data (refers to Table S1).

| Assumptions of stand population and sampling | | Before improving | | | After improving | | |
|---|---|---|---|---|---|---|---|
| | | α | β | e | α | β | e |
| *Homogeneous stands. All stands have same wood density $\rho=0.59$ (t m$^{-3}$) without any biomass measurement errors.* | | | | | | | |
| Regression results are not affected by plot numbers | Average (μ) | 1.79 | 0.87 | 0.0 | 1.79 | 0.87 | 0.0 |
| | SD (σ) | 0.0 | 0.0 | - | 0.0 | 0.0 | - |
| *Heterogeneous stands. Wood densities ($\rho$) are randomly distributed for each stand; measurement errors occur.* | | | | | | | |
| 20 plots | Average (μ) | 1.99 | 0.86 | 0.16 | 1.83 | 0.87 | 0.08 |
| | SD (σ) | 0.385 | 0.038 | - | 0.131 | 0.014 | - |
| 100 plots | Average (μ) | 1.92 | 0.87 | 0.08 | 1.88 | 0.88 | 0.06 |
| | SD (σ) | 0.124 | 0.013 | - | 0.06 | 0.006 | - |

Note: e is residual error (M t) compared between true value (5.18 M t in Table S1) and modeled value.

Fig. 1. A diagram of volume-to-biomass relationship analysis borrowing a simple concept from black-box system identification. The dashed boxes are summed as black-boxes, which could be separated by model postulate, parameter identification, and model validation. e.g. a single module A could include separately two sub-modules expressed by parametric equations.

Fig. 2. An example (*Eucalyptus* and other fast-growing trees in China) of the regression for eqn 1 and eqn 2. The correlation scatters of total biomass vs. stem biomass (blue and top left part) and stem biomass vs. stem volume (green and bottom right part). The top left parts illustrate the regression for eqn 1, and bottom right parts do this for eqn 2. A restricted zone is designed ranging from the lower-bound of $B_t$ (shows a proportion as 80% stem and 20% other parts of the tree) to upper-bound of $B_s$ (shows maximum wood density of 0.7). The samples of $B_s$ vs. $B_t$ are less than the ones of $B_s$ vs. V, because some plots were not measured for roots. Note that the restricted zone may change for different species or forest types in realistic forests. There are higher probabilities near the regression curve than away from the regression curve.

Fig. 3. Spatial distribution of all plots measured across of global forest map (Gong et al., 2013 and Yu et al., 2013 )

Fig. 4. The correlation scatters of 30 tree species and forest types across the world. The scatters show

stem biomass vs. total biomass vs. (curves in top left parts) and stem volume vs. biomass (strait-lines in bottom right parts). The top left parts illustrate the regression for eqn 1, and bottom right parts do this for eqn 2. CD means coefficient of determination; n is the numbers of plots. Every scatter has two red lines that warn the lower limit for eqn 1 and upper limit for eqn 2. Note that these red lines are set up tentatively for approximate estimates of restricted zone that may change for different species and types. The number of each scatter corresponds to the one listed in Table 1. Three variables (volume, stem and total biomass) were measured for most plots. The samples of $B_s$ vs. $B_t$ are less than the ones of V vs. $B_s$, because some plots were not measured for roots. Alternatively, the scatters having fewer samples of V vs. $B_s$ mean that some plots lack measured volumes. The details of data refer to footnotes of Table 1.

Fig. 5. Comparison between two estimations before and after improving the parameters based simulated data. Parameter ($\alpha$ and $\beta$) and residual distributions resulted from 500 simulations (N=500) of setting field plot. The residual errors are abstractive values between total regional biomasses from dummy data and model estimates, which are calculated using the volume-biomass equation with different parameters. These parameters are determined by two samplings for testing different model behaviors by setting up 20 and 100 plots amongst 10,000 stands (n=20 and n=100, population=10,000). The two samplings are executed in each simulation.

Fig. 6. Total biomass of *Eucalyptus* forests in China during 2008-2012, compared with those computed using different parameters. (0) is the original volume-biomass equation directly regressed according to collected field measurements. Numbers from (1) to (5) represent the biomass estimates using "separate-to-recombine" method with different wood densities, $\rho$=0.56, 0.46, 0.61, 0.59, and 0.54, respectively. The calculations of $\rho$ are shown in Fig. 2.

Fig. 7. The variances of volume-to-biomass relationship embodied by the field plots (collected in the three data sets, referring to the footnote of Table 1) across the world's forests. The upper scatter diagram is for tropical trees, and the lower scatter contains all plots for conifer, broadleaf, and mixed trees. The regression condition for $B_{t\,(B_s)}$ was designed to ensure that stem biomass cannot exceed 70% of maximum total biomass (600 t ha$^{-1}$). The red areas show continuous maps of the probability, which corresponding to the wood densities calculated from Zanne et al.'s dataset.

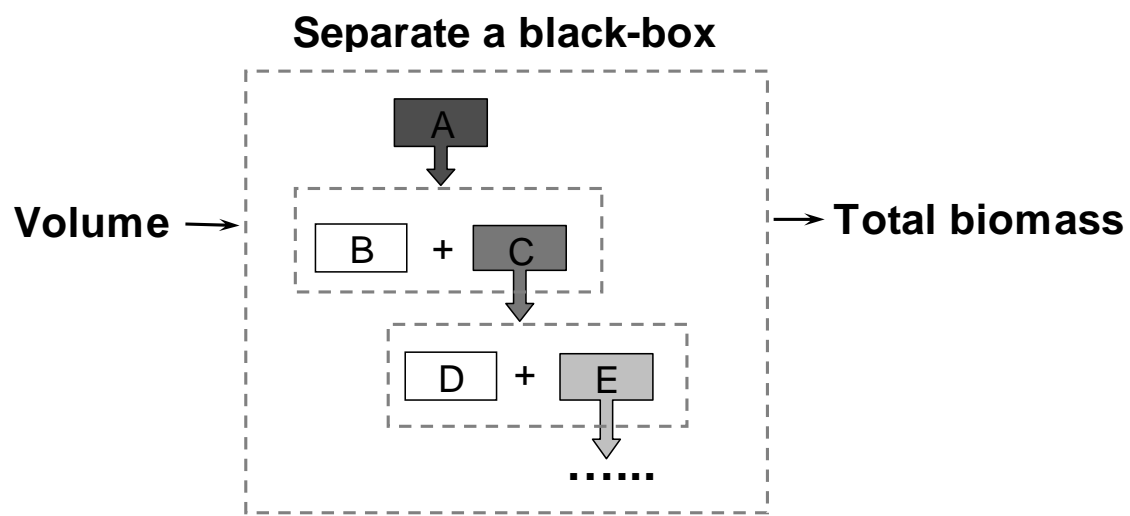

Fig. 1

*Before improving:*

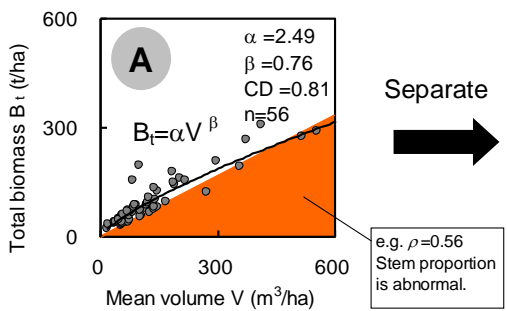

*After improving:*

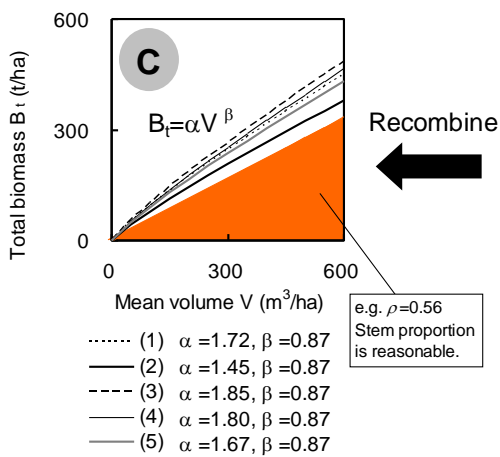

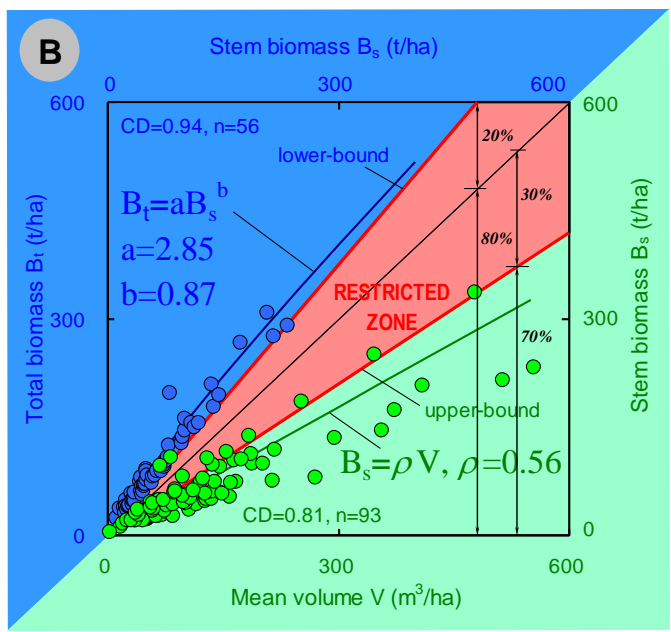

Fig. 2

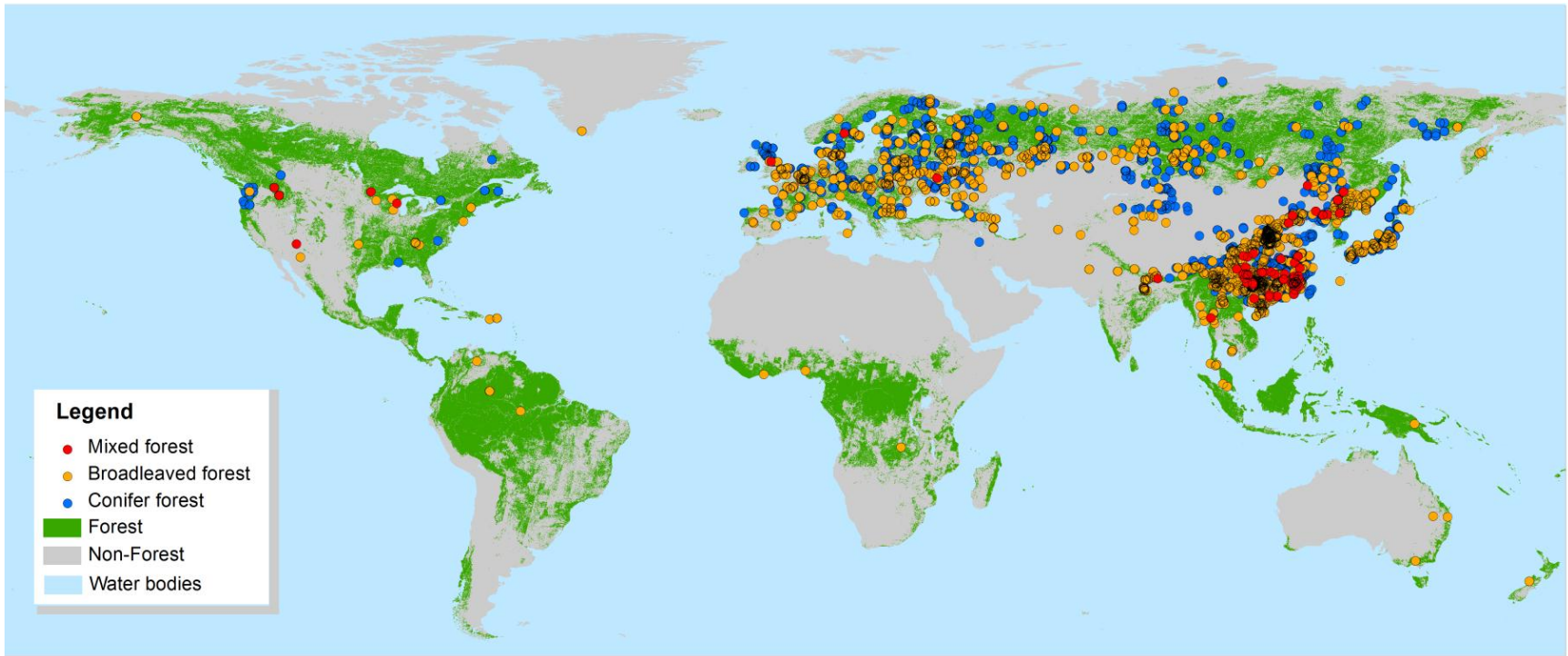

Fig. 3.

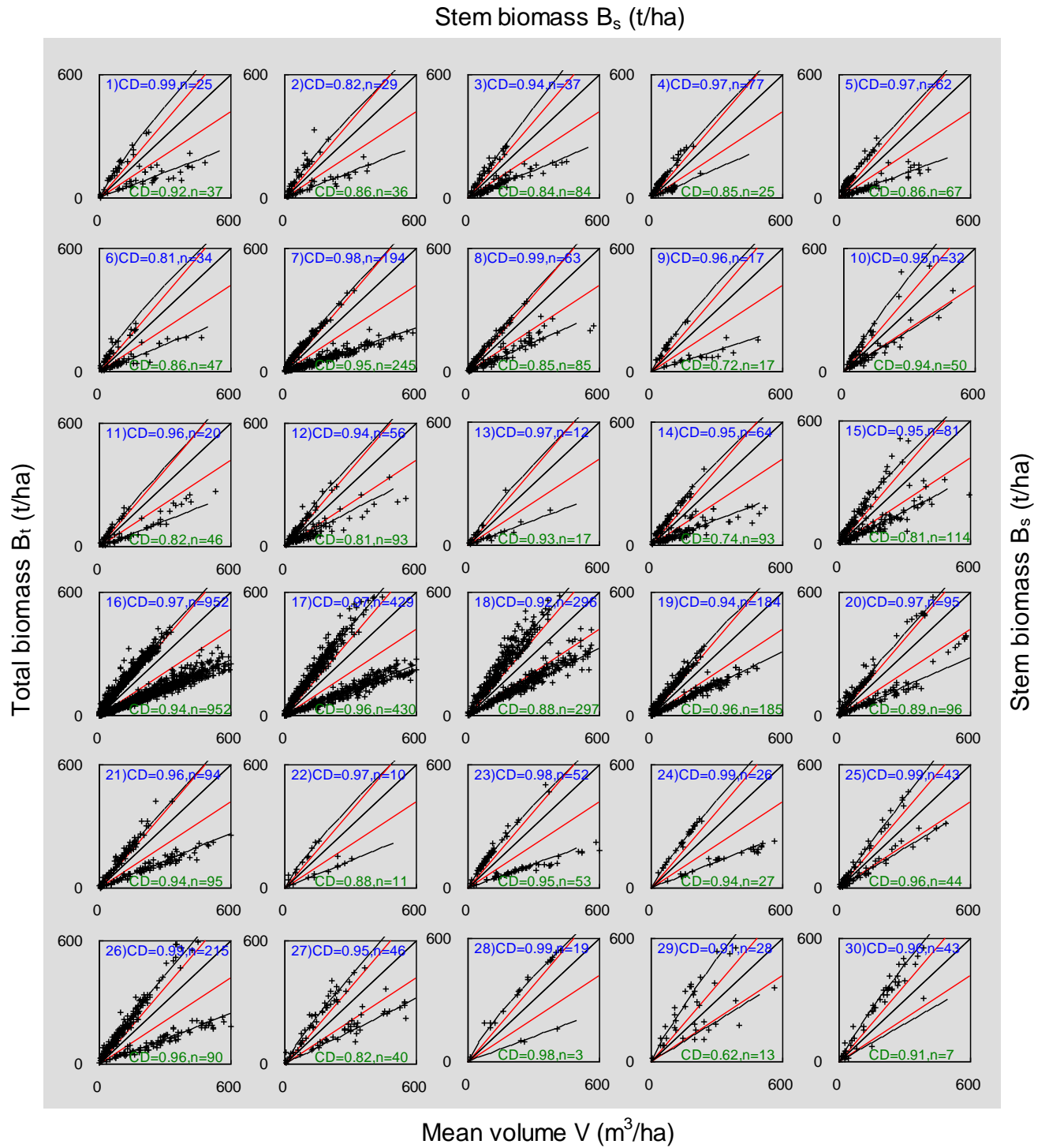

Fig. 4.

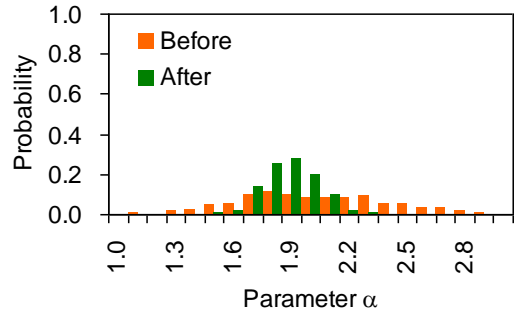
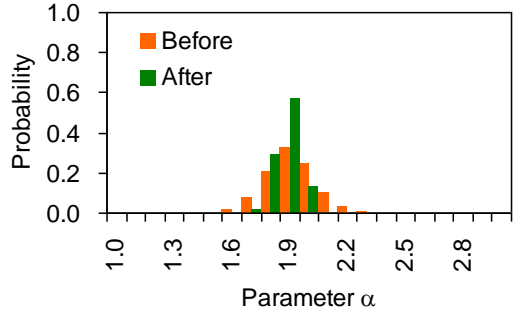
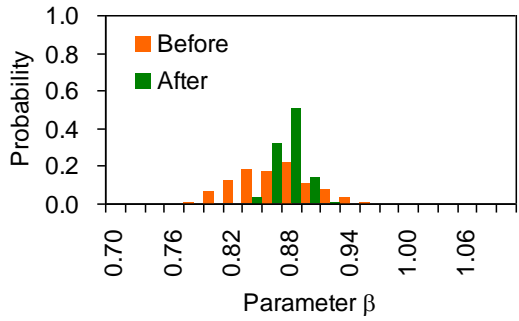
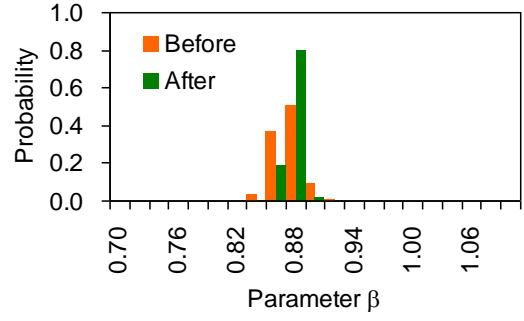
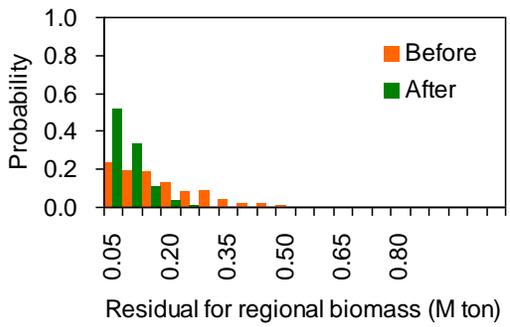
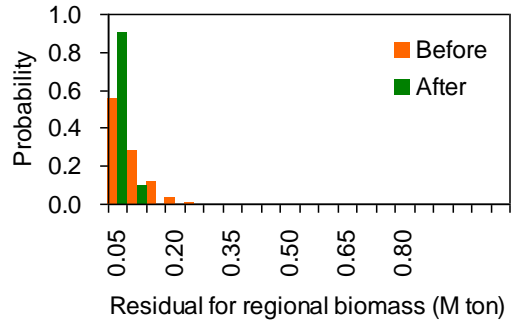

**20-plot-sampling**　　　**100-plot-sampling**

Fig. 5.

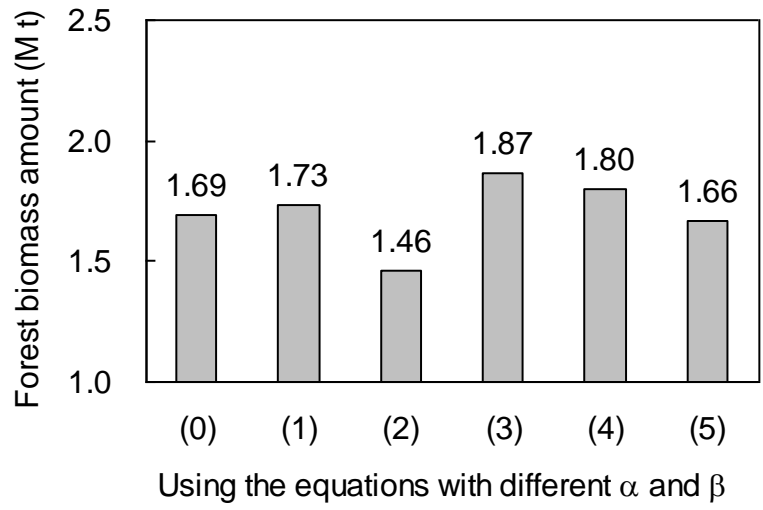

Fig. 6

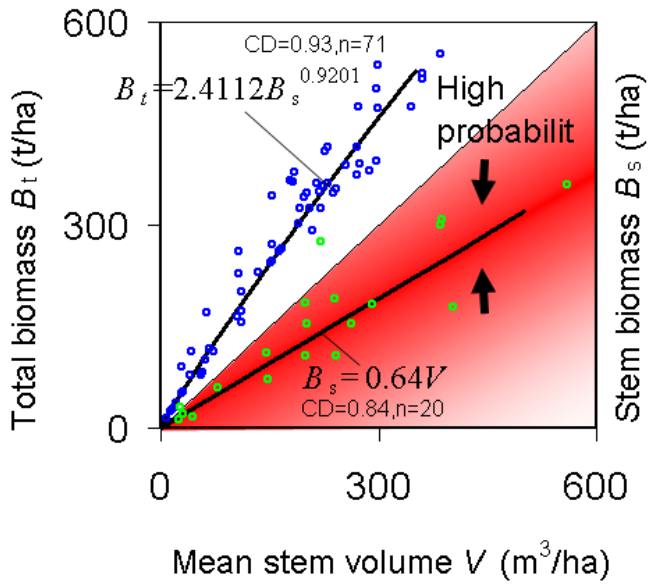
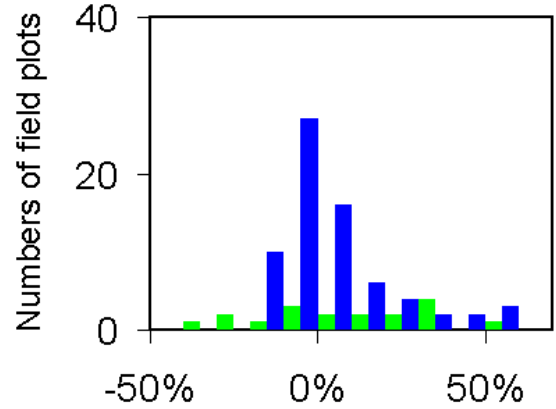
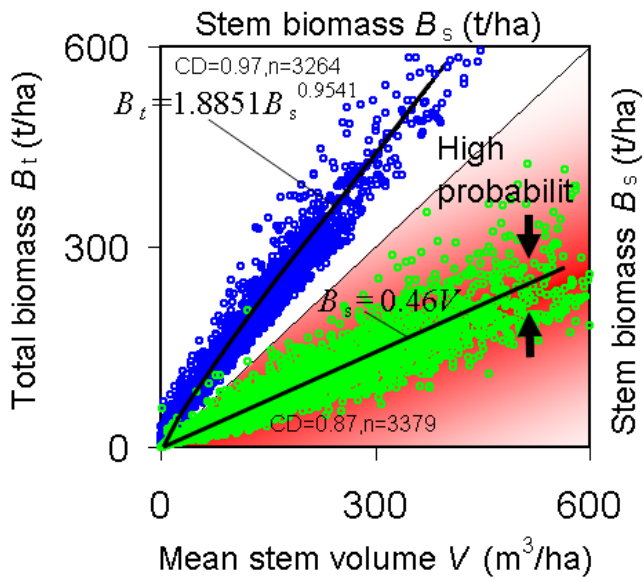
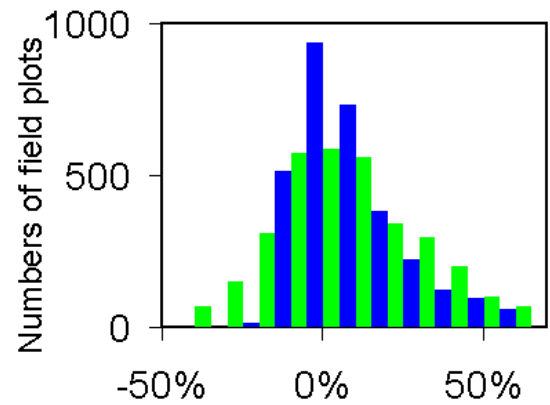

Fig. 7

Table S1. The simulated data set based on two assumptions with even ρ (0.59, t/m³) and randomly distributed ρ for constructing a stand population in a large region (10,000 stands, assuming *Eucalyptus*). The amounts of 5.19 and 5.18 million tons are two true values for comparing estimates using different volume-biomass equations.

| Stand no. | Area (ha) | V (m³ha⁻¹) | Homogeneous stands | | | | Heterogeneous stands | | | |
|---|---|---|---|---|---|---|---|---|---|---|
| | | | ρ (t m⁻³) | Bs (t ha⁻¹) | Bt (t ha⁻¹) | Stand Bt (t) | ρ (t m⁻³) | Bs (t ha⁻¹) | Bt (t ha⁻¹) | Stand Bt (t) |
| 1 | 4.9 | 43.5 | 0.59 | 25.4 | 47.5 | 233.5 | 0.49 | 21.4 | 42.5 | 208.9 |
| 2 | 2.9 | 42.2 | 0.59 | 24.7 | 50.1 | 147.1 | 0.48 | 20.3 | 38.8 | 114.0 |
| 3 | 3.3 | 580.7 | 0.59 | 339.7 | 391.9 | 1280.2 | 0.57 | 331.4 | 447.7 | 1462.6 |
| 4 | 3.8 | 324.2 | 0.59 | 189.6 | 279.1 | 1053.8 | 0.52 | 168.9 | 272.7 | 1029.3 |
| 5 | 3.2 | 64.9 | 0.59 | 38.0 | 72.4 | 233.3 | 0.57 | 37.0 | 67.7 | 218.3 |
| 6 | 1.6 | 96.8 | 0.59 | 56.6 | 98.1 | 159.4 | 0.61 | 59.4 | 103.9 | 168.8 |
| 7 | 3.0 | 48.2 | 0.59 | 28.2 | 46.4 | 140.2 | 0.68 | 32.7 | 55.4 | 167.4 |
| 8 | 2.5 | 49.2 | 0.59 | 28.8 | 51.1 | 130.0 | 0.66 | 32.6 | 66.7 | 169.7 |
| 9 | 4.5 | 278.9 | 0.59 | 163.1 | 248.7 | 1116.9 | 0.63 | 175.1 | 256.3 | 1150.9 |
| 10 | 4.1 | 180.1 | 0.59 | 105.4 | 155.9 | 637.0 | 0.50 | 90.3 | 147.2 | 601.5 |
| … | … | … | … | … | … | … | … | … | … | … |
| 9999 | 4.2 | 168.9 | 0.59 | 98.8 | 158.3 | 661.6 | 0.57 | 95.8 | 143.6 | 600.1 |
| 10000 | 5.0 | 401.0 | 0.59 | 234.6 | 328.0 | 1638.3 | 0.53 | 213.4 | 320.1 | 1599.1 |
| Total or average | 30,127 | 197.8 | 0.59 | | | 5.19 (M t) | 0.59 | | | 5.18 (M t) |

Note: Stand area ranges from 1 to 5 ha, and mean volume ranges from 10 to 600 m³ ha⁻¹. ρ ranges within 0.59 (±5%).

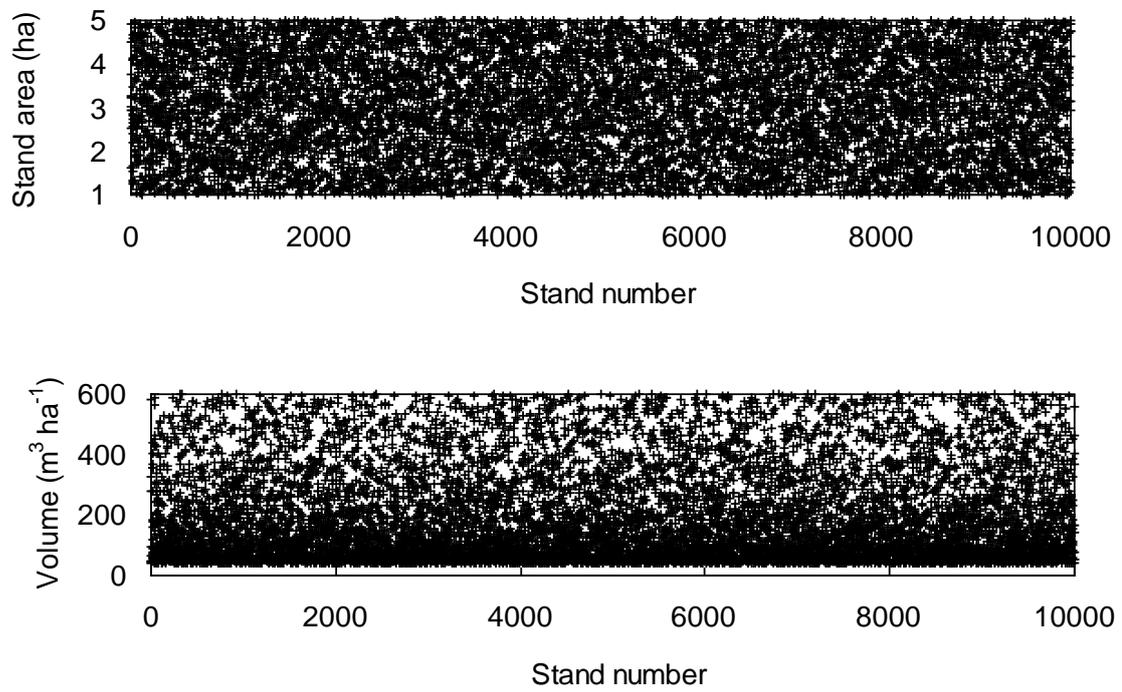

Fig. S1. Simulated stands as a forest population for the model testing experiment. The stand number corresponds to the number in Table S1.

Table S2. Comparison for the ratios of stem to total biomass between different species

| No. | Species or types | a | b | \multicolumn{7}{c|}{Stem biomass (t ha$^{-1}$)} | Ave. † |
| --- | --- | --- | --- | --- | --- | --- | --- | --- | --- | --- | --- |
| | | | | 50 | 100 | 200 | 300 | 400 | 500 | 600 | |
| | | | | \multicolumn{7}{c|}{Ratio of stem to total biomass (R)*} | |
| 1 | *Abies* and *Picea* | 3.32 | 0.86 | 0.52 | 0.57 | 0.63 | 0.67 | 0.70 | 0.72 | 0.74 | 0.65 |
| 2 | *Cupressus* | 3.20 | 0.85 | 0.56 | 0.62 | 0.69 | 0.74 | 0.77 | 0.79 | 0.82 | 0.71 |
| 3 | *Larix* | 1.91 | 0.95 | 0.64 | 0.66 | 0.68 | 0.70 | 0.71 | 0.71 | 0.72 | 0.69 |
| 4 | *Pinus tabulaeformis* | 2.90 | 0.86 | 0.60 | 0.66 | 0.72 | 0.77 | 0.80 | 0.82 | 0.84 | 0.74 |
| 5 | *Pinus koraiensis* and other temperate pines | 3.40 | 0.84 | 0.55 | 0.61 | 0.69 | 0.73 | 0.77 | 0.79 | 0.82 | 0.71 |
| 6 | *Pinus yunnanensis* and other subtropical pines | 4.50 | 0.80 | 0.49 | 0.56 | 0.64 | 0.70 | 0.74 | 0.77 | 0.80 | 0.67 |
| 7 | *Cunninghamia lanceolata* | 2.52 | 0.89 | 0.61 | 0.66 | 0.71 | 0.74 | 0.77 | 0.79 | 0.80 | 0.73 |
| 8 | *Pinus massoniana* | 1.96 | 0.93 | 0.67 | 0.70 | 0.74 | 0.76 | 0.78 | 0.79 | 0.80 | 0.75 |
| 9 | Other conifer trees | 3.80 | 0.83 | 0.51 | 0.58 | 0.65 | 0.69 | 0.73 | 0.76 | 0.78 | 0.67 |
| 10 | Oaks and other deciduous trees | 3.15 | 0.87 | 0.53 | 0.58 | 0.63 | 0.67 | 0.69 | 0.71 | 0.73 | 0.65 |
| 11 | *Populus* and *Betula* | 2.07 | 0.92 | 0.66 | 0.70 | 0.74 | 0.76 | 0.78 | 0.79 | 0.81 | 0.75 |
| 12 | *Eucalyptus* and other fast-growing trees | 2.85 | 0.87 | 0.58 | 0.64 | 0.70 | 0.74 | 0.76 | 0.79 | 0.81 | 0.72 |
| 13 | Soft broadleaved trees | 2.78 | 0.87 | 0.60 | 0.65 | 0.72 | 0.76 | 0.78 | 0.81 | 0.83 | 0.73 |
| 14 | Mixed conifer and deciduous forests | 4.28 | 0.80 | 0.51 | 0.59 | 0.67 | 0.73 | 0.77 | 0.81 | 0.84 | 0.70 |
| 15 | Other hard broadleaved trees | 4.27 | 0.80 | 0.51 | 0.59 | 0.68 | 0.73 | 0.78 | 0.81 | 0.84 | 0.71 |
| 16 | *Pinus* | 2.57 | 0.89 | 0.60 | 0.65 | 0.70 | 0.73 | 0.75 | 0.77 | 0.79 | 0.71 |
| 17 | *Abies* and *Picea* | 2.98 | 0.87 | 0.56 | 0.61 | 0.67 | 0.70 | 0.73 | 0.75 | 0.77 | 0.69 |
| 18 | *Fagus, Acer, Carpinus* and *Quercus* | 2.90 | 0.88 | 0.55 | 0.60 | 0.65 | 0.68 | 0.71 | 0.73 | 0.74 | 0.67 |
| 19 | *Betula* | 3.14 | 0.85 | 0.57 | 0.64 | 0.71 | 0.75 | 0.78 | 0.81 | 0.83 | 0.73 |
| 20 | *Larix* | 2.22 | 0.91 | 0.64 | 0.68 | 0.73 | 0.75 | 0.77 | 0.79 | 0.80 | 0.74 |
| 21 | *Alnus* and *Populus* | 2.13 | 0.91 | 0.67 | 0.71 | 0.76 | 0.78 | 0.81 | 0.82 | 0.83 | 0.77 |
| 22 | *Tilia* | 3.42 | 0.84 | 0.55 | 0.61 | 0.68 | 0.73 | 0.76 | 0.79 | 0.81 | 0.71 |
| 23 | *Castanopsis, Cryptomeria,* and *Pseudotsuga* | 3.39 | 0.84 | 0.55 | 0.62 | 0.69 | 0.73 | 0.77 | 0.80 | 0.82 | 0.71 |
| 24 | *Chamaecyparis obtusa* | 3.23 | 0.86 | 0.54 | 0.59 | 0.65 | 0.69 | 0.72 | 0.74 | 0.76 | 0.67 |
| 25 | *Eucalyptus* and other fast-growing trees | 1.88 | 0.96 | 0.62 | 0.64 | 0.66 | 0.67 | 0.68 | 0.68 | 0.69 | 0.66 |
| 26 | Conifer | 2.59 | 0.90 | 0.57 | 0.61 | 0.66 | 0.68 | 0.70 | 0.72 | 0.73 | 0.67 |
| 27 | Broadleaved | 2.90 | 0.88 | 0.55 | 0.60 | 0.65 | 0.68 | 0.71 | 0.73 | 0.74 | 0.67 |
| 28 | Mixed | 4.04 | 0.81 | 0.52 | 0.59 | 0.68 | 0.73 | 0.77 | 0.81 | 0.83 | 0.71 |
| 29 | Tropical | 3.71 | 0.86 | 0.47 | 0.51 | 0.57 | 0.60 | 0.62 | 0.64 | 0.66 | 0.58 |
| 30 | Tropical | 3.30 | 0.87 | 0.50 | 0.55 | 0.60 | 0.64 | 0.66 | 0.68 | 0.70 | 0.62 |

Note: *Stem ratios (R) of total biomass were calculated with eqn 3 $\Rightarrow R=B_s/B_t=B_s/(aB_s^b)= B_s^{(1-b)}a^{-1}$. † Most stem ratios are close to 0.7 within +/-10%.